\documentclass[]{ptephy_v1}
\usepackage{amsmath}
\usepackage{color}
\usepackage{bm}
\allowdisplaybreaks[3]

\usepackage[normalem]{ulem}  
\renewcommand\sout{\bgroup \color{red} \ULdepth=-.5ex \ULset}

\begin{document}

\title{Asymptotic behavior of Nambu-Bethe-Salpeter wave functions for scalar systems with a bound state}

\author{\name{Shinya~Gongyo}{1} and \name{Sinya~Aoki}{2}}
\address{${}^1$\affil{1}{Theoretical Research Division, Nishina Center, RIKEN, Saitama 351-0198, Japan \email{shinya.gongyo@riken.jp}}
\\
${}^2$\affil{2}{Yukawa Institute for Theoretical Physics, Kyoto University, Kyoto 606-8502, Japan \email{saoki@yukawa.kyoto-u.ac.jp }
}}

\date{\today}

\begin{abstract}
We study the asymptotic behaviors of the Nambu-Bethe-Salpeter (NBS) wave functions,
which are important for the HAL QCD potential method to extract hadron interactions,
in the case that a bound state exists in the system.
We consider the complex scalar particles, two of which lead to the formation of a bound state. 
In the case of the two-body system, we show that the NBS wave functions for the bound state as well as scattering states 
in the asymptotic region behave like the wave functions in quantum mechanics, which carry the information of the binding energy as well as the scattering phase shift. 
This analysis theoretically establishes under some conditions that  the HAL QCD potential 
can correctly reproduce not only the scattering phase shift but also the binding energy.
 As an extension of the analysis, we also study 
the asymptotic behaviors of all possible NBS wave functions in the case of the three-body systems, two of which can form a bound states.
\end{abstract}


\maketitle

\section{Introduction}
\label{Introduction}
  Lattice quantum chromodynamics (QCD) is a successful non-perturbative method to study hadron physics from 
  the underlying degrees of freedom, i.e. quarks and gluons.
  Masses of the single stable hadrons obtained from lattice QCD show good agreement with the experimental results, 
  and even hadron interactions have been recently explored in lattice QCD.
  Using the Nambu-Bethe-Salpeter (NBS) wave function, linked to the S-matrix in QCD
   \cite{Luscher:1990ux, Lin:2001ek, Aoki:2005uf, Ishizuka:2009bx,
   Ishii:2006ec, Aoki:2009ji, Aoki:2012tk, 
Carbonell:2016ekx,Aoki:2013cra}, 
 the hadron interactions have been investigated mainly by two methods: the finite volume method \cite{Luscher:1990ux} and 
  the HAL QCD potential method \cite{ Ishii:2006ec,Aoki:2009ji,Aoki:2012tk}. Theoretically the two methods in principle give same
  results of the scattering phase shift between two hadrons, while in practice they sometimes show different numerical
   results for two baryon systems, whose  origin has been clarified recently in Refs.~\cite{Iritani:2016jie,Iritani:2017rlk}.   
  
 The first method relies on L\"uscher's finite volume formula \cite{Luscher:1990ux}  that relates energies of two hadrons on finite volume  to the phase shifts in infinite volume 
  by utilizing the NBS wave function in the asymptotic region. In practice, energies of the two hadron system are extracted from the temporal correlation of the NBS wave function summed over spatial coordinates and are transformed into the corresponding phase shifts via the finite volume formula. 
  The second method utilizes the NBS wave function in
  non-asymptotic (interacting) region, and extracts the non-local but energy-independent potentials from the space and time dependences of the NBS wave function.
  Physical observables such as phase shifts and binding energies are then calculated by solving the Schr\"odinger equation in infinite volume using the obtained potentials,
  since  the asymptotic behavior of the NBS wave function is related to the $T$-matrix  element
and thus to the phase shifts \cite{Aoki:2013cra}. 
 In practice, the non-local potential is given by the form of the derivative expansion,
 which is truncated by the first few orders \cite{Iritani:2018zbt}. 
This method has been successfully applied to a wide range of two (or three) hadron systems at heavy pion masses 
  \cite{Nemura:2008sp, Inoue:2010hs, Inoue:2010es, Murano:2011nz, Doi:2011gq,Inoue:2011ai,  Murano:2013xxa, Kurth:2013tua, Ikeda:2013vwa, Etminan:2014tya, Yamada:2015cra, Sasaki:2015ifa, Ikeda:2016zwx, Miyamoto:2017tjs, Kawai:2017goq,Ikeda:2017mee} 
  as well as at the nearly physical mass \cite{Gongyo:2017fjb,Sasaki:2017ysy,Ishii:2017xud, Doi:2017cfx, Nemura:2017bbw,Doi:2017zov,Nemura:2017vjc}. 
 
While  a relation of the asymptotic behaviors of the NBS wave functions to the scattering phase shift 
(or more generally the $S$-matrix) is important for both methods, in particular, for the HAL QCD potential method, 
the theoretical arguments for the relation are rather limited: 
Two-body relativistic systems without bound states have been discussed in several different ways
\cite{Lin:2001ek,Aoki:2005uf,Ishizuka:2009bx,Aoki:2010ry}, while
$n$-body non-relativistic systems without bound states have been considered in Ref.~\cite{Aoki:2013cra}, using the Lippmann-Schwinger equation. 

 In these systems, its Hilbert space is of course expanded only by scattering states, so that the asymptotic states are composed of also only the scattering states.
 If the system contains bound states,
on the other hand, the Hilbert space is expanded by  bound states
as well as the scattering states. This situation has never been considered
in previous works and will be discussed in this paper.

The aim of this paper is to relate the asymptotic behaviors of NBS wave functions in scalar systems
to their phase shifts and binding energy in the presence of one bound state.
We apply the Lippmann-Schwinger approach in Ref.~\cite{Aoki:2013cra} to the two- and three-body systems
with a bound state. 
In the approach, we split the Hamiltonian into a free part which reproduces all energy spectrum and an interacting part. 
The free part includes not only scattering states but also a bound state, 
since both appear as asymptotic states and thus there are no reason to exclude the latter.
We first consider the Lippman-Schwinger equations for the two-body scalar system with 
a bound state in Sec.\ref{Lippmann-Schwinger_equation_for_two_scalar_fields}, and then
derive the asymptotic behaviors of the corresponding NBS wave functions in 
Sec.\ref{Asymptotic_states_for_two-body_scalar_system_with_a_bound_state}. 
 We then generalize our analysis  to the three-body scalar system with a bound state:
 We consider the three-body Lippmann-Schwinger equations in 
 Sec.\ref{Lippmann-Schwinger_equation_in_nonrelativistic_three-body_scalar_system_with_a_bound_state_in_two-body_sub_system}, and derive the asymptotic behaviors 
 of the corresponding NBS wave functions in Sec.\ref{Asymptotic_behaviors_of_NBS_wave_functions_for_three-body_nonrelativistic_scalar_system_with_a_bound_state}. 

\section{Lippmann-Schwinger equation for two scalar fields}
\label{Lippmann-Schwinger_equation_for_two_scalar_fields}
 Let us first consider a Hamiltonian $H$ composed of two complex scalar fields with the same (physical) mass $m_a=m_b=m$ (denoted by $\phi_a$ and $\phi_b$) whose interaction leads to an $S$-wave bound state 
 with the (physical) mass $m_B (< 2m)$. In this section, we derive the Lippmann-Schwinger equation in the two-body system, following the definition and notation in Refs. 
 \cite{Aoki:2013cra} and \cite{weinberg2005quantum}. The Hamiltonian is divided into two terms, a free Hamiltonian $H_0$ and an interaction $V$,
 \begin{align}
 H =H_0 +V,
 \end{align}
 where a free eigenstate $\left|\alpha\right>_0$ and  an asymptotic in-state $ \left|\alpha \right>_{\mathrm{in}}$
 satisfy
 \begin{equation}
 H_0 \left|\alpha \right>_0 = E_\alpha\left|\alpha \right>_0, \qquad
 H \left|\alpha \right>_{\mathrm{in}} = E_\alpha\left|\alpha \right>_{\mathrm{in}}
\end{equation}
for the same energy $E_\alpha$.
In order to deal with the bound state in this description,  
we include the scalar field $\phi_B$ corresponding to the bound state composed of $\phi_a$ and $\phi_b$
in the Hamiltonian as
\begin{align}
 H_0 = \int d^3x \left[\sum_{i=a,b}\frac{1}{2}\left(\pi_i ^2 + \left|\phi_i\right|^2 + m^2|\phi_i|^2\right) +\frac{1}{2}\left(\pi_B ^2 + \left|\phi_B\right|^2 + m_B^2|\phi|^2\right)\right],
\end{align}
with $\pi_i (i=a,b)$ and $\pi_B$ conjugate momenta for $\phi_i(i=a,b)$ and $\phi_B$, respectively. 
The Heisenberg operator for  the scalar fields at $t=0$ can be expressed in terms of 
the creation operator of the free anti-particle and annihilation operator of the free particle as
\begin{align}
\phi_i(\mathbf{x}_i,0)=\int \frac{d^3k_i}{\sqrt{(2\pi)^32E_{\mathbf{k}_i}}}\left[a_i(\mathbf{k}_i)e^{i\mathbf{k}_i \cdot \mathbf{x}_i}+b_i^\dagger (\mathbf{k}_i)e^{-i\mathbf{k}_i \cdot \mathbf{x}_i}\right], \quad
 i=a,b, B .
\label{eq:expansion}
\end{align}
with $E_{\mathbf{k}_i}=\sqrt{m_i^2+(\mathbf{k}_i)^2}$. Note that this form fixed at $t=0$
does not hold generally at $t\not= 0$ if $V\not=0$.  

In general, eigen-states $\vert \alpha\rangle_{0,{\rm in}}$ contain, $n_a$ ($n_{\bar a}$) $\phi_a$-particles (anti-particles),
$n_b$ ($n_{\bar b}$) $\phi_b$-particles (anti-particles), and $n_B$ ($n_{\bar B}$) $\phi_B$-bound states (anti-bound states).
In this section, to consider two-body scattering, we focus on states  with $n_a=n_b=1$ and $n_B=1$, denoted as
\begin{eqnarray}
\vert \mathbf{k}^a, \mathbf{k}^b\rangle_{0,{\rm in}},  \qquad  \vert \mathbf{k}^B \rangle_{0,{\rm in}},
\end{eqnarray}
whose eigen-energies are given by
\begin{eqnarray}
E_{\mathbf{k}^a, \mathbf{k}^b} &=&\sqrt{(k^a)^2 + m^2} + \sqrt{(k^b)^2 + m^2},\qquad
E_{\mathbf{k}^B} = \sqrt{(k^B)^2 + m_B^2} 
\end{eqnarray}
with $k^i = \vert \mathbf{k}^i\vert $ ( $i=a,b,B$).
Explicitly, we can write
\begin{eqnarray}
\vert \mathbf{k}^a, \mathbf{k}^b\rangle_{0} &=& a_a^\dagger (\mathbf{k}^a) a_b^\dagger (\mathbf{k}^b)  \vert 0 \rangle_0, \qquad
\vert \mathbf{k}^B \rangle_{0} =   a_B^\dagger (\mathbf{k}^B)  \vert 0 \rangle_0.
\end{eqnarray}

The Lippmann-Schwinger equation formally relates the in-state $\left|\alpha\right>_{\mathrm{in}}$ to
the free-particle state $\left|\alpha\right>_{0}$ as
\begin{align}
\left|\alpha\right>_{\mathrm{in}}= \left|\alpha\right>_{0} + \left(E_\alpha -H_0 +i\epsilon\right)^{-1}V\left| \alpha\right>_{\mathrm{in}} .
\end{align}
By inserting a complete set of free-particle states into the second term, the equation reduces to
\begin{align}
\left|\alpha\right>_{\mathrm{in}}= \left|\alpha\right>_{0} + \frac{1}{2\pi}\int d\beta \frac{\left|\beta\right>_0
T_{\beta; \alpha}}{E_\alpha -E_\beta +i\epsilon},
~~~\frac{T_{\beta;\alpha}}{2\pi}\equiv
{}_0\left<\beta\right|V\left| \alpha\right>_{\mathrm{in}},
\end{align}
where $\int d\beta$ represents both summation over all bound states and integration over all scattering states.
Note that unlike the system without a bound state discussed in Ref.\cite{Aoki:2013cra}, the complete set of free particle states includes
the bound states.

Let us consider the Lippmann-Schwinger equation for the  two particle (scattering) state 
$ \left|\mathbf{k}^a,\mathbf{k}^b \right>_{\mathrm{in}}$ as
\begin{align}
\left|\mathbf{k}^a,\mathbf{k}^b \right>_{\mathrm{in}} = \left|\mathbf{k}^a,\mathbf{k}^b \right>_{\mathrm{0}} 
+\frac{1}{2\pi}\int d^3 q^a\, d^3q^b\, \frac{\left|\mathbf{q}^a,\mathbf{q}^b\right>_0 T_{\mathbf{q}^a,\mathbf{q}^b ;\mathbf{k}^a,\mathbf{k}^b}}{E_{\mathbf{k}^a,\mathbf{k}^b}
-E_{\mathbf{q}^a,\mathbf{q}^b}+ i\epsilon} 
+\frac{1}{2\pi}\int d^3 q^B\frac{\left| \mathbf{q}^B\right>_0 T_{\mathbf{q}^B ;\mathbf{k}^a,\mathbf{k}^b}}{E_{\mathbf{k}^a,\mathbf{k}^b}-E_{\mathbf{q}^B}+ i\epsilon} ,
\end{align}
where the third term appears since the bound state has the same quantum number of the two-particle state with $n_a=n_b=1$. 
We therefore need to consider the equation for the bound state as
\begin{align}
\left| \mathbf{k}^B \right>_{\mathrm{in}} = \left| \mathbf{k}^B\right>_{\mathrm{0}}  
+\frac{1}{2\pi}\int d^3 q^a\,d^3q^b\,\frac{\left|\mathbf{q}^a,\mathbf{q}^b\right>_0 T_{\mathbf{q}^a,\mathbf{q}^b ;\mathbf{k}^B}}{E_{\mathbf{k}^B}-E_{\mathbf{q}^a,\mathbf{q}^b}+ i\epsilon} ,
\end{align}
where the LSZ reduction formula tells us that $T_{\mathbf{q}^B,\mathbf{k}^B}$ vanishes. 
Let us also remind readers that $T_{\mathbf{q}^a,\mathbf{q}^b ;\mathbf{k}^a,\mathbf{k}^b}$ has a pole corresponding to the bound state at an imaginary momentum.

For simplicity we consider the center-of-mass frame, which implies $\mathbf{k}\equiv \mathbf{k}_a=-\mathbf{k}_b$ 
for the scattering state and $\mathbf{k}_B=\mathbf{0}$ for the bound state. 
By using the momentum conservation, the equations can be written as
\begin{align}
\left|\mathbf{k},-\mathbf{k}\right>_{\mathrm{in}}
&= \left|\mathbf{k},-\mathbf{k}\right>_{\mathrm{0}} 
+\frac{1}{2\pi}\frac{\left|\mathbf{q}_B=\mathbf{0}\right>_0 T_{1-2}(\mathbf{0};\mathbf{k})}{E_k-m_B} 
+\frac{1}{2\pi}\int d^3 q\frac{\left|\mathbf{q},-\mathbf{q}\right>_0 T_{2-2}(\mathbf{q};\mathbf{k})}
{E_k-E_q+ i\epsilon} , 
\label{eq:LSeq} \\
\left| \mathbf{k}^B=\mathbf{0}\right>_{\mathrm{in}}
&=\left| \mathbf{k}^B=\mathbf{0}\right>_{\mathrm{0}}  
+\frac{1}{2\pi}\int d^3 q\frac{\left|\mathbf{q},-\mathbf{q}\right>_0T_{2-1}(\mathbf{q};\mathbf{0}) }{m_B-E_q}  \label{eq:LSeq_for_bound}
\end{align}
with $E_{k}=2\sqrt{k^2+m^2}$ ($E_{q}=2\sqrt{q^2+m^2}$), where we have defined the  half off-shell $T$-matrices as 
\begin{align}
T_{\mathbf{q}_B ;\mathbf{k},-\mathbf{k}}&=\delta(\mathbf{q}_B)T_{1-2}(\mathbf{0};\mathbf{k}), \notag \\
T_{\mathbf{q}_a, \mathbf{q}_b ;\mathbf{k},-\mathbf{k}} &=\delta(\mathbf{q}_a+\mathbf{q}_b)T_{2-2}(\mathbf{q};\mathbf{k}) \notag \\
T_{\mathbf{q}_a,\mathbf{q}_b ;\mathbf{k}^B=\mathbf{0}}&=\delta(\mathbf{q}_a+\mathbf{q}_b)T_{2-1}(\mathbf{q};\mathbf{0}),
\end{align}
with $\mathbf{q}=\mathbf{q}_a=-\mathbf{q}_b$.
We have removed some of "$i\epsilon$" if the corresponding denominator does not lead to any real poles.
As mentioned before, $T_{2-2}(\mathbf{q};\mathbf{k})$ has a pole at $\mathbf{k}^2=  \mathbf{q}^2
= m_B^2/4 - m^2 < 0$.

\section{Asymptotic behaviors of the NBS wave functions for two-body scalar system with a bound state}
\label{Asymptotic_states_for_two-body_scalar_system_with_a_bound_state}
In this section, we derive the asymptotic behavior of the equal-time NBS wave functions for two scalar fields, $\phi_a$ and $\phi_b$, in the center of mass system as
\begin{align}
\Psi^{\mathbf{k}}_{ab} (\mathbf{r})&= {}_{\mathrm{in}}\left< 0\right|\phi_a(\mathbf{x_a},0)\phi_b(\mathbf{x_b},0)\left|\mathbf{k},-\mathbf{k}\right>_{\mathrm{in}}.\notag \\
\Psi^B_{ab} (\mathbf{r})&= {}_{\mathrm{in}}\left< 0\right|\phi_a(\mathbf{x_a},0)\phi_b(\mathbf{x_b},0)\left|\mathbf{k}^B=\mathbf{0}\right>_{\mathrm{in}}.\label{eq:NBS}
\end{align}
with $\mathbf{r}=\mathbf{x}_a-\mathbf{x}_b$. By substituting Eq.(\ref{eq:LSeq}) and Eq.(\ref{eq:LSeq_for_bound}) into Eq.(\ref{eq:NBS}), we obtain
\begin{align}
\Psi^{\mathbf{k}}_{ab} (\mathbf{r})&
={}_{\mathrm{in}}\left< 0\right|\phi_a(\mathbf{x_a},0)\phi_b(\mathbf{x_b},0)\left|\mathbf{k},-\mathbf{k}\right>_{\mathrm{0}}
 +\frac{1}{2\pi}\frac{{}_{\mathrm{in}}\left< 0\right|\phi_a(\mathbf{x_a},0)\phi_b(\mathbf{x_b},0)\left|\mathbf{q}^B=\mathbf{0}\right>_0 
 T_{1-2}(\mathbf{0};\mathbf{k})}{E_k-m_B} \notag \\
& +\frac{1}{2\pi}\int d^3 q\, \frac{{}_{\mathrm{in}}\left< 0\right|\phi_a(\mathbf{x_a},0)\phi_b(\mathbf{x_b},0)\left|\mathbf{q},-\mathbf{q}\right>_0 
T_{2-2}(\mathbf{q};\mathbf{k})}{E_k-E_q+ i\epsilon}. \label{eq:NBS2} \\
\Psi^B_{ab} (\mathbf{r})&
={}_{\mathrm{in}}\left< 0\right|\phi_a(\mathbf{x_a},0)\phi_b(\mathbf{x_b},0)\left|\mathbf{k}^B=\mathbf{0}\right>_{\mathrm{0}} \notag \\
&  +\frac{1}{2\pi}\int d^3\, q\frac{{}_{\mathrm{in}}\left< 0\right|\phi_a(\mathbf{x_a},0)\phi_b(\mathbf{x_b},0)\left|\mathbf{q},-\mathbf{q}\right>_0 
T_{2-1}(\mathbf{q};\mathbf{0})}{m_B-E_q}. \label{eq:NBS2_bound}
\end{align}
 
As shown in Appendix \ref{L-Svacuum_in-state}, 
${}_{\mathrm{in}}\left< 0\right|\phi_a(\mathbf{x_a},0)\phi_b(\mathbf{x_b},0)\left|\mathbf{q}^B=0\right>_0$ is exponentially suppressed at large $\vert \mathbf{r}\vert$, 
while ${}_{\mathrm{in}}\left< 0\right|\phi_a(\mathbf{x_a},0)\phi_b(\mathbf{x_b},0)\left|\mathbf{q},-\mathbf{q}\right>_0 $ behaves as 
\begin{align}
{}_{\mathrm{in}}\left< 0\right|\phi_a(\mathbf{x_a},0)\phi_b(\mathbf{x_b},0)\left|\mathbf{q},-\mathbf{q}\right>_0 
\simeq \frac{1}{Z(\mathbf{q})}{}_{0}\left< 0\right|\phi_a(\mathbf{x_a},0)\phi_b(\mathbf{x_b},0)\left|\mathbf{q},-\mathbf{q}\right>_0. 
\end{align}

 Consequently, in the asymptotic region, $\left|\mathbf{x}^a-\mathbf{x}^b\right| \gg 1$, the NBS wave functions reduce to 
 \begin{align}
 \Psi^{\mathbf{k}}_{ab} (\mathbf{r})&
\simeq \frac{1}{Z(\mathbf{k})}{}_{\mathrm{0}}\left< 0\right|\phi_a(\mathbf{x_a},0)\phi_b(\mathbf{x_b},0)\left|\mathbf{k},-\mathbf{k}\right>_{\mathrm{0}} \notag \\
 & +\frac{1}{2\pi}\int d^3 q \frac{1}{Z(\mathbf{q})}\frac{{}_{0}\left< 0\right|\phi_a(\mathbf{x_a},0)\phi_b(\mathbf{x_b},0)\left|\mathbf{q},-\mathbf{q}\right>_0 
T_{2-2}(\mathbf{q};\mathbf{k})}{E_k-E_q+ i\epsilon}, \notag \\
&=\frac{1}{E_k Z(\mathbf{k})} \left[\frac{e^{i\mathbf{k}\cdot\mathbf{r}}}{(2\pi)^3}
+\frac{1}{2\pi}\int\frac{d^3q}{(2\pi)^3}\frac{Z(\mathbf{k})E_k}{Z(\mathbf{q})E_q}\frac{e^{i\mathbf{q}\cdot
\mathbf{r}}T_{2-2}(\mathbf{q};\mathbf{k})}{E_k-E_q+i\epsilon}\right] \\
 \Psi^{B}_{ab} (\mathbf{r})&
= \frac{1}{2\pi}\int d^3 q \frac{1}{Z(\mathbf{q})}\frac{{}_{0}\left< 0\right|\phi_a(\mathbf{x_a},0)\phi_b(\mathbf{x_b},0)\left|\mathbf{q},-\mathbf{q}\right>_0 
T_{2-1}(\mathbf{q};\mathbf{0})}{m_B-E_q} \notag \\
&=\frac{1}{2\pi}\int\frac{d^3q}{(2\pi)^3}\frac{1}{Z(\mathbf{q})E_q}\frac{e^{i\mathbf{q}\cdot
\mathbf{r}}T_{2-1}(\mathbf{q};\mathbf{0})}{m_B-E_q} .
 \end{align}
The rotational invariance implies that $Z(\mathbf{q}) = Z_q$ with $q=\vert \mathbf{q}\vert$, and we can write 
the spherical expansions as
\begin{align}
e^{i\mathbf{q}\cdot \mathbf{r}} &= 4\pi \sum_{l,m}i^lj_l(qr)Y_{lm}(\Omega _{\mathbf{r}})Y^{\ast}_{lm}(\Omega_{\mathbf{q}}), \\
T_{2-2}(\mathbf{q};\mathbf{k})&=\sum_{l,m}T_l^{2-2}(q,k)Y_{lm}(\Omega_{\mathbf{q}})Y^{\ast}_{lm}(\Omega_{\mathbf{k}}), \\
T_{2-1}(\mathbf{q};\mathbf{0})&= T_0^{2-1}(q)Y_{00}(\Omega_{\mathbf{q}}) , \\
\Psi^{\mathbf{k}}_{ab} (\mathbf{r})&= \sum_{l,m}i^l\Psi^{lm}_{ab}(r,k)Y_{lm}(\Omega _{\mathbf{r}})Y^{\ast}_{lm}(\Omega_{\mathbf{k}}), \\
\Psi^{B}_{ab} (\mathbf{r})&= \Psi^{B,00}_{ab}(r)Y_{00}(\Omega _{\mathbf{r}}),
\end{align}
After performing the integration over $\Omega_{\mathbf{q}}$, we obtain
\begin{align}
\Psi ^{lm}_{ab}(r,k) &= \frac{4\pi}{(2\pi)^3E_{k} Z_{k}} \left[j_l(kr)+\int_0^{\infty}\frac{q^2dq}{2\pi }
\frac{Z_kE_k}{Z_q E_q}\frac{j_l(qr)T_l^{2-2}(q,k)}{E_k-E_q+i\epsilon}\right], \label{eq:NBS_sp}\\
\Psi^{B,00}_{ab}(r) &= \frac{4\pi }{(2\pi)^3E_{k}Z_{k}}\int_0^{\infty}\frac{q^2dq}{2\pi}
\frac{Z_kE_k}{Z_q E_q}\frac{j_0(qr)T_0^{2-1}(q)}{m_B-E_q}.\label{eq:NBS_B_sp}
\end{align}

If $k^2$ is below the 4-particle threshold ($2\sqrt{k^2+m^2} < 4m$),
the half off-shell $T$-matrix $T_l^{2S-2S}(q,k)$ does not have any poles or cuts on the real $q$ axis.
Then the integration in Eq.~(\ref{eq:NBS_sp}) can be performed at large $r$ \cite{Aoki:2005uf,Ishizuka:2009bx,Aoki:2009ji,Aoki:2010ry},
by picking up the contribution from the pole at $E_q=E_k+i\epsilon$ as \footnote{The condition for $T_l^{2S-2S}(q,k)$ at $q=0$ assumed in Refs.~\cite{Aoki:2005uf,Aoki:2010ry} is found to be   unnecessary.}
\begin{align}
\Psi ^{lm}_{ab}(r,k) &\simeq \frac{4\pi}{(2\pi)^3E_{k} Z_{k}} \left[j_l(kr)-\frac{kE_k}{8}
\left[n_l(kr) + ij_l(kr)\right]T_l^{2-2}(k,k)\right],
\end{align}
where the contributions from the singularities in the upper half-plane become exponentially small in the asymptotically large $r$ region \cite{Aoki:2010ry}. 
Using the relation between the $T$-matrix and the phase of the $S$-matrix in Appendix \ref{On-shell_T-matrix_and_its_parametrization}, $\Psi ^{lm}_{ab}(r,k)$ is expressed 
as
\begin{align}
\Psi ^{lm}_{ab}(r,k) &\simeq \frac{4\pi}{(2\pi)^3E_{k}Z_{k}} \left[j_l(kr)+
\left[n_l(kr) + ij_l(kr)\right]e^{i\delta_l^{2-2}(k)}\sin \delta_l^{2-2}(k).\right], \notag \\
&\simeq \frac{4\pi}{(2\pi)^3E_{k}Z_k}\frac{e^{i\delta_l^{2-2}(k)}}{kr}
\sin\left(kr -l\pi/2 +\delta_l^{2-2}(k)\right),
\end{align}
which shows that  the NBS wave function encodes the information of the scattering phase shift in its asymptotic behavior as if it were the wave function of the quantum mechanics \cite{Aoki:2005uf,Ishizuka:2009bx,Aoki:2010ry}, so that the potential defined from the NBS wave functions can reproduce the scattering phases shift of QCD \cite{Aoki:2009ji,Aoki:2012tk,Aoki:2013cra} even if the bound state exists in this channel.

The integral in Eq.(\ref{eq:NBS_B_sp}), on the other hand, 
is evaluated at large $r$ as
\begin{eqnarray}
\Psi^{B, 00}_{ab}(r) &\simeq& \frac{ - T_0^{2-1}(i E_B)}{16 \pi^2 Z_{iE_B}}\frac{ e^{- \kappa_B r}}{r} +
\mbox{other exponentially damping contributions} ,
\end{eqnarray}
with $\kappa_B =\sqrt{m^2 - m_B^2/4}$.
In the above, other exponentially damping contributions come from poles or cuts in $Z_q E_q$ or the half-off shell $T$-matrix $T_0^{2-1}(q)$ in the upper half plane of the complex $q$,
which are expected to of the order of the typical mass scale of the system such as $m$.
Therefore   $\Psi^{B, 00}_{ab}(r)$ is dominated by the first term as long as
$\kappa_B$ is smaller than the typical mass scale, which is $\Lambda_{\rm QCD}$ or $m_\pi$ in the case of QCD.
Therefore, the potential constructed from the NBS wave functions including  $\Psi^{B~00}_{ab}(r)$ can reproduce 
the binding energy of the bound state.
 This result ensures that the time dependent HAL QCD method \cite{HALQCD:2012aa} works to extract the potential 
 correctly from the correlation function,  $ \langle 0| \phi_a (\mathbf{x}_a,t) \phi_b (\mathbf{x}_b, t)   \overline{\mathcal{J}} (0)|0\rangle$
 with a source operator $\overline{\mathcal{J}} (0)$, which couples to the bound state together with scattering states.
This is a main result of this paper.

\section{Lippmann-Schwinger equation in the nonrelativistic three-body scalar system with a bound state in a two-body sub system}
\label{Lippmann-Schwinger_equation_in_nonrelativistic_three-body_scalar_system_with_a_bound_state_in_two-body_sub_system}
As an extension of the analysis in the previous sections, we consider 
the system of three complex scalar fields, $\phi_i~ \left(i=a,b,c\right)$ with the same mass $m$,
where the interaction between $\phi_a $ and $\phi_b $ leads to one bound state denoted 
by $\phi_B$ with a mass $m_B (< 2m)$, and the other two-body interactions do not lead to any other bound states.
We examine how interactions between the fundamental scalar particles and the bound particle are encoded in their NBS wave functions.

In this and the following sections, we consider 
\begin{eqnarray}
\left|\mathbf{k}^a, \mathbf{k}^b, \mathbf{k}^c\right>_0  
&=& a^\dagger_a(\mathbf{k}^a) a^\dagger_b(\mathbf{k}^b) a^\dagger_c(\mathbf{k}^c) 
\vert 0 \rangle_0, \quad
\left|\mathbf{k}^c, \mathbf{k}^B\right>_0  
=  a^\dagger_c(\mathbf{k}^c) a^\dagger_B(\mathbf{k}^B)
\vert 0 \rangle_0, 
\end{eqnarray}
which are eigenstates of the free Hamiltonian $H_0$ coupled with each other by the interaction,
where the corresponding energies  are given by
\begin{eqnarray}
E_{\mathbf{k}^a, \mathbf{k}^b, \mathbf{k}^c} &=&
\sqrt{(k^a)^2+m^2}+ \sqrt{(k^b)^2+m^2}+\sqrt{(k^c)^2+m^2}, \notag \\
E_{\mathbf{k}^c,\mathbf{k}^B} &=& \sqrt{(k^c)^2+m^2}+ \sqrt{(k^B)^2+m_B^2}.
\end{eqnarray}
As before, the scalar fields in the Heisenberg representation at $t=0$ can be expressed in terms of the creation and annihilation operators as
\begin{align}
\phi_i(\mathbf{x}_i,0)=\int \frac{d^3k_i}{\sqrt{(2\pi)^32E_{\mathbf{k}^i}}}\left[a_i(\mathbf{k}_i)e^{i\mathbf{k}_i \cdot \mathbf{x}_i}+b_i^\dagger (\mathbf{k}_i)e^{-i\mathbf{k}_i \cdot \mathbf{x}_i}\right], \qquad i=a,b,c, B .
 \label{eq:expansion3}
\end{align}
 
For the three-body system,
it is convenient to introduce the modified Jacobi coordinates \cite{Aoki:2013cra} as,
\begin{align}
\mathbf{\tilde{x}}_i =\sqrt{ \frac{i}{i+1}}\mathbf{x}_i^J,~~\mathbf{\tilde{k}}_i = \sqrt{{\frac{i+1}{i}}}\mathbf{k}_i^J, ~~ (i=1,2)
\end{align}
where $\mathbf{x}_i^J$ and $\mathbf{k}_i^J$ are the standard Jacobi coordinates,
\begin{align}
&\mathbf{x}_1^J = \mathbf{x}_1 -\mathbf{x}_2,~ \mathbf{x}_2^J = \frac{1}{2}\left(\mathbf{x}_1 +\mathbf{x}_2\right)-\mathbf{x}_3, ~\mathbf{x}_3^J 
= \frac{1}{3}\left(\mathbf{x}_1 +\mathbf{x}_2+\mathbf{x}_3\right), \notag \\
&\mathbf{k}_1^J = \frac{1}{2}(\mathbf{k}_1 -\mathbf{k}_2),~ \mathbf{k}_2^J 
= \frac{2}{3}\left\{\frac{1}{2}\left(\mathbf{k}_1 +\mathbf{k}_2\right)-\mathbf{k}_3\right\}, ~\mathbf{k}_3^J = \mathbf{k}_1 +\mathbf{k}_2+\mathbf{k}_3.
\end{align}
In the center of mass frame, we have
\begin{align}
\sum_{i=1}^3\mathbf{k}_i\cdot \mathbf{x}_i =\sum_{i=1}^{2}\mathbf{\tilde{k}}_i\cdot \mathbf{\tilde{x}}_i,~~E_{\mathbf{k}_1,\mathbf{k}_2,\mathbf{k}_3}\simeq 3m 
+ \sum_{i=1}^{3}\frac{\mathbf{k}_i^2}{2m} =3m+\frac{1}{2m}\sum_{i=1}^{2}(\mathbf{\tilde{k}}_i)^2\equiv E_{\mathbf{\tilde{k}}_1,\mathbf{\tilde{k}}_2},
\end{align}
where  the nonrelativistic approximation is used for the energy.
In these coordinates, three-body noninteracting state and in-state in the center of mass frame can be parametrized as
\begin{align}
\left|\mathbf{k}^a,\mathbf{k}^b,\mathbf{k}^c\right>_{0,\mathrm{in}} &=|\mathbf{\tilde{k}}^a,\mathbf{\tilde{k}}^b\left. \right>_{0,\mathrm{in}}.
\end{align}

The Lippmann-Schwinger equations in the center of mass frame for 
$ \left|\mathbf{\tilde{k}}^a,\mathbf{\tilde{k}}^b \right>_{\mathrm{in}} $
and $ \left|\mathbf{k}^c,\mathbf{k}^B=-\mathbf{k}^c\right>_{\mathrm{in}}$ are given within the nonrelativistic approximation as
\begin{align}
|\mathbf{\tilde{k}}^a,\mathbf{\tilde{k}}^b\left. \right>_{\mathrm{in}} 
&\simeq |\mathbf{\tilde{k}}^a,\mathbf{\tilde{k}}^b\left. \right>_{\mathrm{0}} 
+ \frac{1}{2\pi}\frac{1}{3^{3/2}}\int d^3 \tilde{q}^ad^3\tilde{q}^b\frac{\left|\mathbf{\tilde{q}}^a,\mathbf{\tilde{q}}^b\right>_0
 T_{3-3}(\mathbf{\tilde{q}}^a,\mathbf{\tilde{q}}^b;\mathbf{\tilde{k}}^a,\mathbf{\tilde{k}}^b)}
{E_{\mathbf{\tilde{k}}^a,\mathbf{\tilde{k}}^b}-E_{\mathbf{\tilde{q}}^a,\mathbf{\tilde{q}}^b}+ i\epsilon} \notag \\
&+
\frac{1}{2\pi}\int d^3 q^c\frac{\left|\mathbf{q}^c,\mathbf{q}^B=-\mathbf{q}^c\right>_0T_{2-3}(\mathbf{q}^c;\mathbf{\tilde{k}}^a,\mathbf{\tilde{k}}^b)}
{E_{\mathbf{\tilde{k}}^a,\mathbf{\tilde{k}}^b}-E_{\mathbf{q}^c}^B + i\epsilon}
 \label{eq:LS_three}\\
\left|\mathbf{k}^c ,\mathbf{k}^B=-\mathbf{k}^c \right>_{\mathrm{in}} 
&\simeq \left|\mathbf{k}^c,\mathbf{k}^B =-\mathbf{k}^c \right>_{\mathrm{0}} 
+\frac{1}{2\pi}\int d^3 q^c\frac{\left|\mathbf{q}^c,\mathbf{q}^B=-\mathbf{q}^c\right>_0 T_{2-2}(\mathbf{q}^c;\mathbf{k}^c)}{E_{\mathbf{k}^c}^B -E_{\mathbf{q}^c}^B + i\epsilon} \notag \\
 &+\frac{1}{2\pi}\frac{1}{3^{3/2}}\int d^3 \tilde{q}^ad^3\tilde{q}^b\frac{\left|\mathbf{\tilde{q}}^a,\mathbf{\tilde{q}}^b\right>_0 T_{3-2}(\mathbf{\tilde{q}}^a,\mathbf{\tilde{q}}^b;\mathbf{k}^c)}
{E_{\mathbf{k}^c}^B-E_{\mathbf{\tilde{q}}^a,\mathbf{\tilde{q}}^b}+ i\epsilon}, \label{eq:LS_one_bound}
\end{align}
 where $E_{\mathbf{k}^c}^B \simeq m_B + m + \frac{(\mathbf{k}^c)^2}{2m_{\rm red}}$ with $1/m_{\rm red} =1/m +1/m_B$, 
 and
\begin{align}
 T_{\mathbf{q}^a,\mathbf{q}^b,\mathbf{q}^c ;\mathbf{k}^a,\mathbf{k}^b,\mathbf{k}^c}&=\delta(\mathbf{q}^a+\mathbf{q}^b+\mathbf{q}^c)T_{3-3}(\mathbf{\tilde{q}}^a,\mathbf{\tilde{q}}^b;\mathbf{\tilde{k}}^a,\mathbf{\tilde{k}}^b),\\
 T_{\mathbf{q}^c\mathbf{q}^B ;\mathbf{k}^a,\mathbf{k}^b,\mathbf{k}^c}&=\delta(\mathbf{q}^c+\mathbf{q}^B )T_{2-3}(\mathbf{q}^c;\mathbf{\tilde{k}}^a,\mathbf{\tilde{k}}^b), \\
 T_{\mathbf{q}^c\mathbf{q}^B ;\mathbf{k}^c,\mathbf{k}^B} &=\delta(\mathbf{q}^c+\mathbf{q}^B )T_{2-2}(\mathbf{q}^c;\mathbf{k}^c), \\
 T_{\mathbf{q}^a,\mathbf{q}^b,\mathbf{q}^c ;\mathbf{k}^c,\mathbf{k}^B}&=\delta(\mathbf{q}^a+\mathbf{q}^b+\mathbf{q}^c)T_{3-2}(\mathbf{\tilde{q}}^a,\mathbf{\tilde{q}}^b;\mathbf{k}^c).
\end{align}

For the latter convenience, by arranging the momenta as $\mathbf{Q}_2 =\mathbf{q}^c,\mathbf{K}_2 =\mathbf{k}^c, 
\mathbf{Q}_3 =(\mathbf{\tilde{q}}^a,
\mathbf{\tilde{q}}^b)$ and $\mathbf{K}_3 =(\mathbf{\tilde{k}}^a,\mathbf{\tilde{k}}^b)$,
we rewrite the above coupled channel $T$-matrix (or $S$-matrix) compactly as
$T_{l-m}(\mathbf{Q}_l; \mathbf{K}_m)$ (or $S_{l-m}(\mathbf{Q}_l; \mathbf{K}_m)$) with $l,m=2,3$.
 
\section{Asymptotic behaviors of NBS wave functions for three-body nonrelativistic scalar system with a bound state}
\label{Asymptotic_behaviors_of_NBS_wave_functions_for_three-body_nonrelativistic_scalar_system_with_a_bound_state}
The NBS wave functions in the coupled channel are compactly written as
\begin{eqnarray}
\Psi(\mathbf{X}_l \vert \mathbf{K}_m) &=& {}_{\rm in}\langle 0\vert
\Phi_l (\mathbf{x}_l) \vert \mathbf{K}_m \rangle_{\rm in}, 
\end{eqnarray}
where $\mathbf{X}_2 \equiv \mathbf{x}_c -\mathbf{x}_B$, 
$\mathbf{X}_3 \equiv (\mathbf{\tilde{x}}_a,  \mathbf{\tilde{x}}_b)$, and
\begin{eqnarray}
\Phi_2(\mathbf{x}_2)  &\equiv& \phi_c(\mathbf{x}_c,0)  \phi_B(\mathbf{x}_B,0), \quad
\Phi_3(\mathbf{x}_3)  \equiv \phi_a(\mathbf{x}_a,0)  \phi_b(\mathbf{x}_b,0)  \phi_c(\mathbf{x}_c,0) ,
\notag \\
\vert \mathbf{K}_2 \rangle_{\rm in} &=& \vert \mathbf{k}^c, \mathbf{k}^B= -\mathbf{k}^c \rangle_{\rm in} , \qquad
\vert \mathbf{K}_3 \rangle_{\rm in} = \vert \mathbf{\tilde{k}}^a, \mathbf{\tilde{k}}^b \rangle_{\rm in} .
\end{eqnarray}

By using the Lippmann-Schwinger equation given in Eq.(\ref{eq:LS_three}) and Eq.(\ref{eq:LS_one_bound}), the NBS wave functions within the nonrelativistic approximation can be written as
\begin{eqnarray}
\Psi(\mathbf{X}_l \vert \mathbf{K}_m) &=&  {}_{\rm in}\langle 0\vert
\Phi_l (\mathbf{x}_l) \vert \mathbf{K}_m \rangle_{\rm 0}
+\frac{1}{2\pi}\sum_{n=2,3} \int [d\mathbf{P}]_n \frac{  {}_{\rm in}\langle 0\vert
\Phi_l (\mathbf{x}_l) \vert \mathbf{P}_n \rangle_{\rm 0} T_{n-m}(\mathbf{P}_n; \mathbf{K}_m)}
{E_{K_m} - E_{P_n} +i\epsilon},
\label{eq:NBS_three}
\end{eqnarray}
where
\begin{eqnarray}
[d\mathbf{P}]_n &\equiv & 3^{(6-3n)/2} d^{(3n-3)} \mathbf{P}_n = 3^{(6-3n)/2} P_n^{3n-4} dP_n \, d\Omega_{\mathbf{P}_n}, \quad P_n = \vert \mathbf{P}_n\vert, \\
E_{P_2} &=& m_B + m + \frac{P_2^2}{2m_{\rm red}}, \quad
E_{P_3} = 3m + \frac{P_3^2}{2m} .
\end{eqnarray}

As shown in Appendix~\ref{L-Svacuum_in-state}, we have
\begin{eqnarray}
 {}_{\rm in}\langle 0\vert \Phi_l (\mathbf{x}_l) \vert \mathbf{K}_m \rangle_{\rm 0}
 &\simeq &\delta_{lm} D_l( \mathbf{K}_l ) e ^{ i  \mathbf{K}_l \cdot\mathbf{X}_l}
 +\delta_{l2}\delta_{m3}D_{23}(\mathbf{K}_3)e^{i \mathbf{K}_2 \cdot\mathbf{X}_2}
\end{eqnarray}
in the asymptotic region, $|\mathbf{x}_i-\mathbf{x}_j| \gg 1$ for $i,j=a,b,c$ or
$|\mathbf{x}_c-\mathbf{x}_B| \gg 1$, 
where
\begin{eqnarray}
D_2( \mathbf{K}_2 ) &=& \frac{1}{Z_B(\mathbf{k}^c) (2\pi)^3 \sqrt{E_{\mathbf{k}^c} E^B_{\mathbf{k}^c}}}, \quad D_3(\mathbf{K}_3 ) = \frac{1}{Z(\mathbf{\tilde{k}}^a,\mathbf{\tilde{k}}^b)}\prod_{j=a,b,c}\frac{1}{\sqrt{(2\pi)^32E_{\mathbf{k}^j}}}, \notag \\
D_{23}(\mathbf{K}_3) &=&
\frac{1}{2\pi}
\frac{
T^{\dagger}_{0-ab\bar{B}}(0;{\mathbf{k}^a,\mathbf{k}^b},\mathbf{k}^c)
}{(2\pi)^3
(4E_{\mathbf{k}^{c}} E^B_{\mathbf{k}^{c}})^{1/2} (-E_{\mathbf{k}^{a},\mathbf{k}^b,\mathbf{q}^B=\mathbf{k}^c}) } ,\notag \\
E_{\mathbf{k}} &\simeq& m + \frac{  \mathbf{k}^2}{2m}, \quad
E^B_{\mathbf{k}} \simeq m_B + \frac{  \mathbf{k}^2}{2m_B}, \notag \\
\mathbf{K}_2 \cdot\mathbf{X}_2 &=& \mathbf{k}_c \cdot(\mathbf{x}_c - \mathbf{x}_B),
\quad \mathbf{K}_3 \cdot\mathbf{X}_3 =\mathbf{\tilde{k}}_a \cdot\mathbf{\tilde{x}}_a + \mathbf{\tilde{k}}_b 
\cdot\mathbf{\tilde{x}}_b \notag .
\end{eqnarray}
Thus the asymptotic behaviors of NBS wave functions reduce to
\begin{eqnarray}
\Psi(\mathbf{X}_l \vert \mathbf{K}_m) &\simeq&
D_l ( \mathbf{K}_l )\left[ \delta_{lm} e ^{ i  \mathbf{K}_l \cdot\mathbf{X}_l}
+ \frac{1}{2\pi} \int [d\mathbf{P}]_l \frac{D_l ( \mathbf{P}_l )}{D_l ( \mathbf{K}_l )}
\frac{  e^{i \mathbf{P}_l \cdot\mathbf{X}_l} T_{l-m}(\mathbf{P}_l; \mathbf{K}_m)}
{E_{K_m} - E_{P_l} +i\epsilon} \right] \notag \\
&+&D_{23}(\mathbf{K}_3)\delta_{l2}\left[\delta_{m3}e^{i\mathbf{K}_2\cdot \mathbf{X}_2}
+\frac{1}{2\pi}\int [d\mathbf{P}]_3 \frac{D_{23} ( \mathbf{P}_3 )}{D_{23} ( \mathbf{K}_3 )}
\frac{  e^{i \mathbf{P}_2 \cdot\mathbf{X}_2} T_{3-m}(\mathbf{P}_3; \mathbf{K}_m)}
{E_{K_m} - E_{P_3} +i\epsilon} \right].
\end{eqnarray}
Note that the third and forth terms appear due to the transition from vacuum to the state with anti-bound particle and the $a$- and $b$-particles. They do not appear in the coupled channel three-body system 
in the absence of bound states, as was studied in
Ref.~\cite{Aoki:2013cra}.
We introduce the hyperspherical expansion as
\begin{eqnarray}
e^{ i  \mathbf{K}_l \cdot\mathbf{X}_l} &=& d_{3l-3} \sum_{[L]_l} i^L j_L^{3l-3}( K_l X_l) Y_{[L]_l}(\Omega_{\mathbf{X}_l}) Y_{[L]_l}^\ast(\Omega_{\mathbf{K}_l}), \\
T_{l-m}(\mathbf{P}_l; \mathbf{K}_m) &=& \sum_{[L]_l, [M]_m} T^{l-m}_{[L]_l, [M]_m}(P_l, K_m)
Y_{[L]_l}(\Omega_{\mathbf{P}_l}) Y_{[M]_m}^\ast(\Omega_{\mathbf{K}_m}), \\
\Psi(\mathbf{X}_l \vert \mathbf{K}_m) &=& \sum_{[L]_l, [M]_m} \Psi_{[L]_l, [M]_m}(X_l, K_m)
Y_{[L]_l}(\Omega_{\mathbf{X}_l}) Y_{[M]_m}^\ast(\Omega_{\mathbf{K}_m}),
\end{eqnarray}
where $X_l = \vert \mathbf{X}_l\vert$, 
\begin{eqnarray}
d_D&=& (D-2)!! \frac{2\pi^{D/2}}{\Gamma\left(\frac{D}{2}\right)},
\end{eqnarray}
and 
$j_L^{D}$ is the hyperspherical Bessel function of the first kind,
defined by
\begin{eqnarray}
j_L^D(x) &=& \frac{\Gamma\left(\frac{D-2}{2}\right) 2^{\frac{D-4}{2}}}{(D-4)!! x^{\frac{D-2}{2}}}
J_{L_D}(x),
\end{eqnarray}
with $L_D = L +\frac{D-2}{2}$ and the Bessel function of the first kind $J_{L_D}(x)$.
Explicitly, they are expressed as
\begin{eqnarray}
j_L^3 (x) &\equiv&\sqrt{\frac{\pi}{2x}}J_{L+1/2}(x)= j_L(x), \quad j_L^6(x) \equiv \frac{J_{L+2}(x)}{x^2},
\end{eqnarray}
where  $j_L(x)$ is the spherical Bessel function of the first kind.
Here $Y_{[L]_l}$ is the $3l-3$ dimensional hyperspherical harmonic function, which satisfies
\begin{eqnarray}
\int d\Omega\, Y_{[L]_l}(\Omega)  Y_{[M]_l}^\ast(\Omega) &=& \delta_{[L]_l,[M]_l},
\end{eqnarray}
and $Y_{[L]_2}$ corresponds to the spherical harmonic function $Y_{lm}$.
Within the non-relativistic approximation, we can write $D_l(\mathbf{K}_l) $ and 
$D_{23}(\mathbf{K}_3)$ as
$D_l(\mathbf{K}_l) \simeq D_l ( K_l), ~D_{23}(\mathbf{K}_3) \simeq D_{23}^2 ( K_2)D_{23}^3 ( K_3)
$ with $K_l =\vert \mathbf{K}_l \vert$. 

The integration over $d\Omega_{\mathbf{P}_l}$  gives
\begin{eqnarray}
\Psi_{[L]_l, [M]_m}(X_l, K_m) &=& d_{3l-3} i^L D_l(K_l) 
\Biggl[ j_L^{3l-3}( K_l X_l) \delta_{[L]_l,[M]_l}\delta_{lm}\notag \\
&+& \frac{3^{(6-3l)/2}}{2\pi}\int dP_l P_l^{3l-4} \frac{D_l(P_l)}{D_l(K_l)}
\frac{j_L^{3l-3}(P_l X_l) T^{l-m}_{[L]_l, [M]_m}(P_l, K_m)}{E_{K_m} -E_{P_l} +i\epsilon}
\Biggr] \notag \\
&+& d_{3} i^L\delta_{l2}D_{23}^3(K_3)\Biggl[   \delta_{m3} J_{[L]_2 [M]_3}(K_3, X_2) \notag \\
&+& \frac{3^{-3/2}}{2\pi}\int dP_3 P_3 ^5\frac{D_{23}^3(P_3)}{D_{23}^3(K_3)}
\frac{\sum_{[N]_3} J_{[L]_2 [N]_3}(P_3, X_2)  T^{3-m}_{[N]_3, [M]_m}(P_3, K_m)}
{E_{K_m} - E_{P_3} +i\epsilon}
\Biggr],~~~~~~~~~~
\end{eqnarray}
where
\begin{align}
J_{[L]_2 [M]_3}(K_3,X_2)=\int d\Omega_{\mathbf{K}_3}D_{23}^2(K_2) j_L^{3}(K_2X_2) 
Y_{[M]_3}(\Omega_{\mathbf{K}_3})Y_{[L]_2}^\ast(\Omega_{\mathbf{K}_2}).
\end{align}
In the nonrelativistic expansion, $D_{23}(\mathbf{K}_3)$ starts from $ O\left(T^\dagger_{0-ab\bar{B}}
/(m^{3/2}m_B^{1/2})\right)$, while $D_{2}(\mathbf{K}_2)$ starts from $ O(1/m)$.
Assuming that $T^\dagger_{0-ab\bar{B}} \simeq O(1)$, 
we therefore drop the third and fourth terms in our analysis hereafter. 

The integration over $P_l$  for the large $X_l$ in the second term
can be performed by picking up the poles inside the closed contour 
 in a complex plane as has been done in Ref. \cite{Aoki:2013cra}, 
\begin{eqnarray}
\Psi_{[L]_l, [M]_m}(X_l, K_m) &\simeq&  i^L D_l(K_l) \frac{ (2\pi)^{\frac{3l-3}{2}}} {(K_l X_l)^{\frac{3l-5}{2}}}
\Biggl[ J_{L_{3l-3}} ( K_l X_l) \delta_{[L]_l,[M]_l}\delta_{l,m}\notag \\
&-& \frac{1}{2}\frac{K_l^{3l-5} m_l }{ 3^{(3l-6)/2}}
\left[ N_{L_{3l-3}}(K_l X_l) + i J_{L_{3l-3}}(K_l X_l) \right]
T^{l-m}_{[L]_l, [M]_m}(K_l, K_m)
\Biggr], ~~
\end{eqnarray} 
where $K_l$ and $K_m$ satisfy $E_{K_l}=E_{K_m} = E$ for a given total energy $E$, and
$N_L(x)$ is the Bessel function of the second kind.
Explicitly,
$K_2$ and $K_3$ are written as $K_2= \sqrt{2m_{\rm red}(E - m_B -m)}$ 
and $K_3= \sqrt{2m (E - 3m)}$.

Using the parametrization of the $T$-matrix derived in Appendix~\ref{On-shell_T-matrix_and_its_parametrization} as
\begin{eqnarray}
 && T_{[L]_l,[M]_m}(K_l, K_m)\nonumber \\
&=&
-2C_l(E) \left\{\sum_{[N]_n} U_{[L]_l,[N]_n} (E) e^{i \delta_{[N]_n}(E)} \sin(  \delta_{[N]_n}(E)) U^\dagger_{[N]_n,[M]_m}(E)\right\} C_m(E)
\end{eqnarray}
with 
\begin{eqnarray}
C_l(E) = \sqrt{\frac{3^{(3l-6)/2}}{ m_l (K_l)^{3l-5}}},
\end{eqnarray}
and the asymptotic behaviors of $J_L(x)$ and $N_L(x)$ for large $\vert x\vert$ as
\begin{eqnarray}
J_L(x) &\simeq & \sqrt{\frac{2}{\pi x}}\sin(x-\Delta_L), \quad
N_L(x) \simeq  \sqrt{\frac{2}{\pi x}}\cos(x-\Delta_L), \qquad
\Delta_L=\frac{2L-1}{2}\pi, 
\end{eqnarray}
we obtain the asymptotic forms related to the phase shifts  as
\begin{eqnarray}
\Psi_{[L]_l, [M]_m}(X_l, K_m) &\simeq& 2 i^L D_l(K_l)  \left(\frac{2\pi}{K_l X_l}\right)^{\frac{3l-4}{2}}
\frac{C_m(E)}{C_l(E)}\sum_{[N]_n} U_{[L]_l,[N]_n}(E) e^{i\delta_{[N]_n}(E)}U^\dagger_{[N]_n, [M]_m} (E) \notag \\
&\times& \sin( K_l X_l -\Delta_{L_{3l-3}} + \delta_{[N]_n}(E) ) ,
\end{eqnarray}
for $l,m=2$ ( $B$, $c$ ) and $l,m=3$ ($a,b,c$).
The final result agrees with the one obtained in the absence of bound states~\cite{Aoki:2013cra}.

\section{Summary and concluding remarks}
 In this paper, we have investigated the asymptotic behaviors of the NBS wave functions in complex scalar systems in the presence of a bound state. 
 We include a bound state in the asymptotic states of the theory and consider
 a coupled system of two scattering particles and one bound state.
 We have shown that
 the asymptotic form for the NBS wave function of the scattering state is related 
 to the phase shifts of $S$-matrix via its unitary, while the asymptotic form for the NBS function of the bound state decays exponentially with its binding energy 
 as long as the binding energy is smaller than other mass scales such as the mass of the scattering particle. 
 This result establishes that the potential extracted  via the time-dependent HAL QCD 
 method \cite{HALQCD:2012aa} from the four-point correlation functions, which couple to the bound state as well as scattering states, can correctly reproduce both binding energy  
and scattering phase shift. This is the main result of this paper.
  
As an extension of our analysis,
 we have considered the three complex scalar systems, two of which can form a bound state.
In addition to the the elastic scattering of the three-particles,
three particles are scattered into one fundamental particle plus one bound state and vice versa. 
Although analysis becomes rather involved in this case, 
the final result in the non-relativistic limit becomes almost identical to the one for the coupled two and three particle systems in Ref.~\cite{Aoki:2013cra}. 

\section*{Acknowledgements}
The authors thank T. Doi, T. Hyodo, and Y. Ikeda for fruitful discussions and useful comments.
S.G. is supported by the Special Postdoctoral Researchers Program of RIKEN.
S. A. is supported in part by the Grant-in-Aid of the Japanese Ministry of Education, Sciences and Technology, Sports and Culture (MEXT) for Scientific Research (No. JP16H03978),  
by a priority issue (Elucidation of the fundamental laws and evolution of the universe) to be tackled by using Post ``K" Computer, 
and by Joint Institute for Computational Fundamental Science (JICFuS).

\appendix

\section{On-shell T-matrix and its  parametrization}
\label{On-shell_T-matrix_and_its_parametrization}
In this appendix, we parametrize the $T$-matrix, where $S=1-iT$,
using the unitarity of the $S$-matrix, $S^\dagger S=1$.
We first discuss a simple two-body system and then consider a three-body system 
with a bound state.

\subsection{Two-body case}
In the center-of-mass frame that ${\mathbf k}^a = -{\mathbf k}^b ={\mathbf k}$,
the $S$-matrix is denoted as
\begin{eqnarray}
{}_0\left<{\mathbf q}^a, {\mathbf q}^b \right| S \left| {\mathbf k}^a={\mathbf k},{\mathbf k}^b=-{\mathbf k}\right>_0 &= 
\delta(E_q-E_k)\delta^{(3)}({\mathbf q}^a+{\mathbf q}^b) 
S_{2-2}({\mathbf q};{\mathbf k}),
\end{eqnarray}
where ${\mathbf q}^a = -{\mathbf q}^b ={\mathbf q}$,
$E_q=2\sqrt{q^2+m^2}$, and $E_k=2\sqrt{k^2+m^2}$ with 
$q=\vert {\mathbf q}\vert$ and $k=\vert{\mathbf k}\vert$. 

The unitarity of the $S$-matrix reads
\begin{eqnarray}
\delta(E_q-E_k)
\int d^3p \delta(E_p-E_k) S_{2-2}({\mathbf q};{\mathbf p}) S_{2-2}^\dagger({\mathbf p};{\mathbf k}) &=& \delta^{(3)}({\mathbf q}-{\mathbf k}). 
\end{eqnarray}
Introducing expansions in terms of the spherical harmonic functions as
\begin{eqnarray}
S_{2-2}({\mathbf q};{\mathbf p})&=& \sum_{l,m} S_l^{2-2}(q;p)Y_{lm}(\Omega_{\mathbf q})
Y_{lm}^\ast(\Omega_{\mathbf p}), \\
\delta^{(3)}({\mathbf q}-{\mathbf k}) &=& \sum_{l,m}\frac{1}{q^2}\delta(q-k)
Y_{lm}(\Omega_{\mathbf{q}})Y^{\ast}_{lm}(\Omega_{\mathbf{k}})
\end{eqnarray}
and
\begin{eqnarray}
\delta(E_p-E_k) &=& \frac{E_q}{4q}\delta(q-k), \qquad
d^3p = p^2 dp d\Omega_{\mathbf{p}}, 
\end{eqnarray}
we obtain
\begin{eqnarray}
\left|S^{2-2}_l(q,q)\right|^2&=&\frac{16}{q^2E_q^2},
\end{eqnarray}
which is solved as 
\begin{eqnarray}
S^{2-2}_l(q,q)&=&\frac{4e^{2i\delta_l^{2-2}(q)}}{qE_q},
\end{eqnarray}
with the phase shift $\delta_l^{2-2}(q)$.
Thus, the $T$-matrix, defined from the $S$-matrix as
$\delta(E_q-E_k) S_{2-2}({\mathbf q}; {\mathbf k}) =\delta({\mathbf q}-{\mathbf k}) - i \delta(E_q-E_k) T_{2-2}({\mathbf q}; {\mathbf k})$ ,
 is expressed as
\begin{eqnarray}
T_{2-2}({\mathbf q};{\mathbf k})&=& \sum_{l,m} T_l^{2-2}(q,q)Y_{lm}(\Omega_{\mathbf q})
Y_{lm}^\ast(\Omega_{\mathbf k}), 
\end{eqnarray}
where  
\begin{eqnarray}
T_l^{2-2} (q,q)=-\frac{8}{qE_q}e^{i\delta_l^{2-2}(q)}\sin \delta_l^{2-2}(q).
\end{eqnarray}

\subsection{Three-body case}
We next consider the scattering of the three-body system, where
not only the elastic scattering but also the bound particle production and its inverse process
occur:
\begin{align}
\phi_a+\phi_b+\phi_c &\rightarrow \phi_a+\phi_b+\phi_c,\  \phi_B+\phi_c, \notag \\
\phi_B+\phi_c &\rightarrow \phi_a+\phi_b+\phi_c,\  \phi_B+\phi_c. \notag 
\end{align}
Denoting the corresponding $S$-matrix as $S_{l-m} (\mathbf{Q}_l;\mathbf{K}_m)$, as
for the $T$-matrix in the main text, the unitarity condition is expressed as
\begin{eqnarray}
&&\delta(E_{Q_l} - E_{K_m}) \sum_{n=2,3}\int d \mathbf{P}_n
\delta(E_{Q_l} - E_{P_n}) S(\mathbf{Q}_l,\mathbf{P}_n)
S^\dagger(\mathbf{P}_n,\mathbf{K}_m)\nonumber \\
& =&\delta_{l,m} 3^{(3l-6)/2}\delta^{(3l-3)}(\mathbf{Q}_l-\mathbf{K}_l),
\end{eqnarray}
where $l,n,m =2,3$ represent a number of particles involved in the state, 
\begin{eqnarray}
\mathbf{P}_2 &=& \mathbf{p}, \quad \mathbf{P}_3 = \mathbf{\tilde{P}}, \quad
E_{P_2} = m_B+ m +\frac{P_2^2}{2m_{\rm red}}, \quad E_{P_3} =3m +\frac{P_3^2}{2m}  ,\\
d \mathbf{P}_n &=& 3^{(3n-6)/2} d^{(3n-3)} \mathbf{P}_n =  3^{(3n-6)/2} (P_n)^{3n-4} dP_n d\Omega_{ \mathbf{P}_n}, \quad P_n =\vert  \mathbf{P}_n\vert.
\end{eqnarray}

Introducing the hyperspherical expansion as\cite{Aoki:2013cra}
\begin{eqnarray}
S(\mathbf{Q}_l; \mathbf{K}_m) &=& \sum_{[L]_l,[M]_m}
S_{[L]_l,[M]_m}(Q_l,K_m) Y_{[L]_l}(\Omega_{\mathbf{Q}_l})
Y_{[M]_m}^\ast(\Omega_{\mathbf{K}_m}), \\
\delta^{(3l-3)}(\mathbf{Q}_l-\mathbf{K}_l) &=& \sum_{[L]_l} \frac{\delta(Q_l - K_l)}{Q^{3l-4}}  Y_{[L]_l}(\Omega_{\mathbf{Q}_l}) Y_{[L]_l}^\ast(\Omega_{\mathbf{K}_l}),
\end{eqnarray}
and using
\begin{eqnarray}
\delta(E_{Q_l} - E) &=&\frac{m_l}{Q_l}\delta(Q_l -Q_l(E)), 
\end{eqnarray}
where $m_2=m_{\rm red}$, $m_3=m$, and $Q_l(E)$ is a solution of $E_{Q_l}=E$,
the unitarity condition for a total energy $E$ reads
\begin{eqnarray}
&&C_l(E)^{-1}\, \sum_{[N]_n} S_{[L]_l,[N]_n}(Q_l(E), Q_n(E))\, C_n(E)^{-1}  \nonumber \\
&\times&C_n(E)^{-1}\, \overline{S_{[M]_m,[N]_n}(Q_m(E), Q_n(E)) }\, C_m(E)^{-1}
=\delta_{[L]_l,[M]_m}, 
\end{eqnarray}
where
\begin{eqnarray}
C_l(E) = \sqrt{\frac{3^{(3l-6)/2}}{ m_l \{Q_l(E)\}^{3l-5}}}.
\end{eqnarray}

Thus we can parametrize the above $S$-matrix as
\begin{eqnarray}
  S_{[L]_l,[M]_m}(Q_l(E), Q_m(E))
&=& C_l(E)\left\{ \sum_{[N]_n}U_{[L]_l,[N]_n} (E) e^{i 2\delta_{[N]_n}(E)} U^\dagger_{[N]_n,[M]_m}(E) \right\} C_m(E),~~~~~~~~
\end{eqnarray}
where $\delta_{[N]_n}(E)$ is the (generalized) scattering phase shift for the channel $[N]_n$ at the energy $E$, and $U(E)$ is the unitary matrix introduced for the diagonalization.

The corresponding $T$-matrix  is then given by
\begin{eqnarray}
 && T_{[L]_l,[M]_m}(Q_l(E), Q_m(E))\nonumber \\
&=& 
-2C_l(E) \left\{\sum_{[N]_n} U_{[L]_l,[N]_n} (E) e^{i \delta_{[N]_n}(E)} \sin(  \delta_{[N]_n}(E)) U^\dagger_{[N]_n,[M]_m}(E)\right\} C_m(E).
\end{eqnarray}

\section{Lippmann-Schwinger equation for the vacuum in-state}
\label{L-Svacuum_in-state}
Using the Lippmann-Schwinger equation for the vacuum in-state, 
\begin{eqnarray}
\left|0\right>_{\mathrm{in}}= \left|0\right>_{0} + \int d\beta \frac{\left|\beta\right>_0T_{\beta; 0}}{ -E_\beta +i\epsilon}, 
\qquad E_0=0,
\label{eq:LSvacuum}
\end{eqnarray}
we evaluate the vacuum contribution to the two-body and three-body NBS wave functions. 
Extensions to the general $n$-body NBS wave functions are straightforward.

\subsection{Two-body case}
As seen in Eq.(\ref{eq:NBS2}) and Eq.(\ref{eq:NBS2_bound}),
(i)$~{}_{\mathrm{in}}\left< 0\right|\phi_a(\mathbf{x_a},0)\phi_b(\mathbf{x_b},0)\left| 
\mathbf{q}^B=0\right>_0$ 
and 
(ii)$~{}_{\mathrm{in}}\left< 0\right|\phi_a(\mathbf{x_a},0)\phi_b(\mathbf{x_b},0)\left|\mathbf{q}^a,\mathbf{q}^b \right>_0$
appear in two-body NBS wave function. 

(i) Using  the Lippmann-Schwinger equation for the vacuum in-state  in Eq.(\ref{eq:LSvacuum}), we have
\begin{eqnarray}
~{}_{\mathrm{in}}\left< 0\right|\phi_a(\mathbf{x_a},0)\phi_b(\mathbf{x_b},0)\left|\mathbf{q}^B=0\right>_0 &=& 
\frac{1}{2\pi}\int d\beta\,  T^{\dagger}_{0;\beta} \frac{  {}_0\left<\beta\right|\phi_a(\mathbf{x_a},0)\phi_b(\mathbf{x_b},0)\left|\mathbf{q}^B=0\right>_0}{ -E_\beta +i\epsilon} , ~~~~~~~
\label{eq:L-S:bound_part}
\end{eqnarray}
since $~{}_{0}\left< 0\right|\phi_a(\mathbf{x_a},0)\phi_b(\mathbf{x_b},0)\left| \mathbf{q}^B=0\right>_0 =0$ from the definition of $\phi_i( \mathbf{x},0)$ in Eq.~(\ref{eq:expansion}).
Non-zero contribution in Eq.(~\ref{eq:L-S:bound_part}), which comes from
\begin{eqnarray}
{}_0\left<\beta\right| = {}_0\left< 0 \right| b_a(\mathbf{k}^a) b_b(\mathbf{k}^b) a_B( \mathbf{k}^B)
\end{eqnarray}
is evaluated as
\begin{eqnarray}
&& ~{}_{\mathrm{in}}\left< 0\right|\phi_a(\mathbf{x}^a,0)\phi_b(\mathbf{x}^b,0)\left|\mathbf{q}^B=0\right>_0 \notag \\
&=& \frac{1}{2\pi}\int d^3k^Bd^3k^ad^3k^b  \,
T^{\dagger}_{0;\mathbf{k}^a,\mathbf{k}^b,\mathbf{k}^B}
\dfrac{
 {}_0\left< 0 \right| b_a(\mathbf{k}^a) b_b(\mathbf{k}^b) a_B( \mathbf{k}^B)
 \phi_a(\mathbf{x}^a,0)\phi_b(\mathbf{x}^b,0)\left|\mathbf{q}^B=0\right>_0}
{ -E_{\mathbf{k}^a,\mathbf{k}^b,\mathbf{k}^B} } \notag \\
&=&\int d^3k^a\dfrac{T^{\dagger}_{0-2} (0;\mathbf{k}^a)\, e^{-i\mathbf{k}^a\cdot (\mathbf{x}^a-\mathbf{x}^b)}}
{(2\pi)^4 2E_{\mathbf{k}^a}(-2E_{\mathbf{k}^a} -m_B)},
\end{eqnarray}
where $T^{\dagger}_{0-2} (0;\mathbf{k}^a)$ is defined as $T^{\dagger}_{0;\mathbf{k}^a,\mathbf{k}^b,\mathbf{k}^B=0} = \delta^{(3)}(\mathbf{k}^a+\mathbf{k}^b)T^{\dagger}_{0-2} (0;\mathbf{k}^a)$. 
Since the integrand in the last line does not have any poles on the real axis of $\vert \mathbf{k}^a\vert$ as in the case of  Sec. \ref{Asymptotic_states_for_two-body_scalar_system_with_a_bound_state}, 
this contribution vanishes exponentially in large $\left|\mathbf{x}^a-\mathbf{x}^b\right|$.
Note that complex poles from the denominator appear at $k^2 = -m^2$ or
on the unphysical sheet of $k^2$ which satisfies $2E_{{\mathbf k}} + m_B=0$.

 (ii) In the same manner, we obtain
\begin{eqnarray}
 ~{}_{\mathrm{in}}\left< 0\right|\phi_a(\mathbf{x_a},0)\phi_b(\mathbf{x_b},0)\left|\mathbf{q}^a,\mathbf{q}^b\right>_0 
 &=& ~{}_{0}\left< 0\right|\phi_a(\mathbf{x_a},0)\phi_b(\mathbf{x_b},0)\left|\mathbf{q}^a,\mathbf{q}^b \right>_0\notag \\
  &+&\frac{1}{2\pi} \int d\beta\, T^{\dagger}_{0;\beta} \dfrac{{}_0\left<\beta\right|\phi_a(\mathbf{x_a},0)\phi_b(\mathbf{x_b},0)\left|\mathbf{q}^a,\mathbf{q}^b\right>_0}{ -E_\beta +i\epsilon}. ~~
 \end{eqnarray}
 As shown in Appendix A of Ref.\cite{Aoki:2013cra},  this reduces to
\begin{eqnarray}
 ~{}_{\mathrm{in}}\left< 0\right|\phi_a(\mathbf{x_a},0)\phi_b(\mathbf{x_b},0)\left|\mathbf{q};-\mathbf{q}\right>_0 &\simeq& \frac{1}{Z(\mathbf{q})}  ~{}_{\mathrm{0}}
  \left< 0\right|\phi_a(\mathbf{x_a},0)\phi_b(\mathbf{x_b},0)\left|\mathbf{q}; -\mathbf{q}\right>_0 
\end{eqnarray}
in the center of mass system,
where $Z(\mathbf{q})$  corresponds to the renormalization factor for the vacuum,
whose explicit form can be found in Ref.\cite{Aoki:2013cra}.

\subsection{Three-body case}
As seen in Eq.(\ref{eq:NBS_three}),
we need to evaluate ${}_{\rm in}\langle 0\vert \Phi_l(\mathbf{x}_l) \vert \mathbf{K}_m\rangle_0$.
Using the Lippmann-Schwinger equation for the vacuum state,
we have
\begin{eqnarray}
{}_{\rm in}\langle 0\vert \Phi_l(\mathbf{x}_l) \vert \mathbf{K}_m\rangle_0
&=& {}_{0}\langle 0\vert \Phi_l(\mathbf{x}_l) \vert \mathbf{K}_m\rangle_0 \delta_{lm}
+ \frac{1}{2\pi} \int d\beta T_{0;\beta}^\dagger \frac{{}_0\left<\beta\right|
\Phi_l(\mathbf{x}_l)
 \left|\mathbf{K}_m\right>_0}{ -E_\beta +i\epsilon} .
  \label{eq:L-S:three_body_bound_part}
\end{eqnarray} 

We first consider the case with $l=m$,
which has already been analyzed
 in Appendix A of Ref.\cite{Aoki:2013cra} for $n=2$ and $n=3$,
and the result is given by
\begin{eqnarray}
{}_{\rm in}\langle 0 \vert \Phi_l(\mathbf{x}_l) \vert \mathbf{K}_l\rangle_0 \simeq
 D_l(\mathbf{K}_l) e^{i \mathbf{K}_l \cdot \mathbf{X}_l} 
\end{eqnarray} 
for the asymptotic region, where
\begin{eqnarray}
D_2(\mathbf{K}_2) &=& \frac{1}{ Z(\mathbf{k}_c) (2\pi)^3\sqrt{4 E_{\mathbf{k}^c} E_{\mathbf{k}^c}}},
\quad 
\frac{1}{Z(\mathbf{k}_c)} = 1+\sum_{i=c,B}\frac{T_{0;i\bar i}^\dagger(0;\mathbf{k}_i, -\mathbf{k}_i)}{- 2\sqrt{k_i^2+m_i^2}}, \\
D_3(\mathbf{K}_3) &=& \frac{1}{Z(\mathbf{\tilde{k}}^a, \mathbf{\tilde{k}}^b)}
\prod_{j=a,b,c} \frac{1}{\sqrt{(2\pi)^3 2E_{\mathbf{k}^j}}},
\quad
\frac{1}{Z(\mathbf{\tilde{k}}^a, \mathbf{\tilde{k}}^b)} =
1+ \sum_{i=a,b,c}\frac{T_{0;i\bar i}^\dagger(0;\mathbf{k}_i, -\mathbf{k}_i)}{- 2\sqrt{k_i^2+m_i^2}} .
~~~~
\end{eqnarray} 
Here $m_{a,b,c}=m$, $\mathbf{k}_c+\mathbf{k}_B =0$ ($\mathbf{k}_b+\mathbf{k}_b+\mathbf{k}_c =0$) for $D_2$ ($D_3$), and $T_{0; i \bar i}$ is the off-shell $T$-matrix from vacuum to a pair of particle-antiparticle with a flavor $i$.

We consider other cases with $l\not=m$. \\
(i) $l=3$ and $m=2$.
Non-zero contributions to ${}_0\left<\beta\right|\phi_a(\mathbf{x_a},0)\phi_b(\mathbf{x_b},0)\phi_c(\mathbf{x_c},0)\left| \mathbf{k}^c,\mathbf{k}^B\right>_0$  come from
\begin{align}
{}_0\left<\beta\right| ={}_0\left<0\right|a_B(\mathbf{q}^B) b_a(\mathbf{q}^{\bar a})b_b(\mathbf{q}^{\bar b})
 ~~\mathrm{or}~~
 {}_0\left<0\right|a_c(\mathbf{q}^c)a_B(\mathbf{q}^B) b_a(\mathbf{q}^{\bar a})b_b(\mathbf{q}^{\bar b})b_c(\mathbf{q}^{\bar c}).
\end{align}
Eq.(\ref{eq:L-S:three_body_bound_part}) thus reduces to
\begin{eqnarray}
&&{}_{\rm in}\langle 0\vert \Phi_l(\mathbf{x}_3) \vert \mathbf{K}_2\rangle_0\notag \\
&=&\frac{1}{2\pi}\int d^3q^{\bar a}\frac{
T^{\dagger}_{0-\bar{a}\bar{b}B}(0;{\mathbf{q}^{\bar a},\mathbf{k}^c-\mathbf{q}^{\bar a},\mathbf{k}^B})
e^{i\mathbf{q}^{\bar a}\cdot (\mathbf{x}_b-\mathbf{x}_a)}e^{i\mathbf{k}^c\cdot (\mathbf{x}_c-\mathbf{x}_b)}
}{\{(2\pi)^3\}^{3/2} 
(8E_{\mathbf{q}^{\bar a}}E_{\mathbf{k}^c-\mathbf{q}^{\bar a}}E_{\mathbf{k}^c})^{1/2}(-E_{\mathbf{q}^{\bar a},\mathbf{k}^c-\mathbf{q}^{\bar a},\mathbf{k}^B}) } \notag \\
&+&\frac{1}{2\pi}\int d^3q^{\bar a}d^3q^{\bar b} 
\frac{
T^{\dagger}_{0-\bar{a}\bar{b}\bar{c}cB}(0;\mathbf{q}^{\bar a},\mathbf{q}^{\bar b}, -\mathbf{q}^{\bar a}-\mathbf{q}^{\bar b}, \mathbf{k}^c, \mathbf{k}^B)
e^{i\mathbf{q}^{\bar a}\cdot (\mathbf{x}_c-\mathbf{x}_a)}e^{i\mathbf{q}^{\bar b}\cdot (\mathbf{x}_c-\mathbf{x}_b)}
}
{\{(2\pi)^3\}^{3/2} (8E_{\mathbf{q}^{\bar a}}E_{\mathbf{q}^{\bar b}}E_{\mathbf{q}^{\bar a}+\mathbf{q}^{\bar b}})^{1/2}(-E_{\mathbf{q}^{\bar a},\mathbf{q}^{\bar b},
-\mathbf{q}^{\bar a}-\mathbf{q}^{\bar b}, \mathbf{k}^c, \mathbf{k}^B}) }
\end{eqnarray}
for $\mathbf{k}^c = -\mathbf{k}^B$.
Since both integrands have no poles for real momenta $\mathbf{q}^{\bar a}$ and/or $\mathbf{q}^{\bar b}$, these terms vanish asymptotically large $\vert \mathbf{x}_{a}-\mathbf{x}_{b}\vert$,
$\vert \mathbf{x}_{a}-\mathbf{x}_{c}\vert$, and $\vert \mathbf{x}_{c}-\mathbf{x}_{b}\vert$.\\

\noindent
 (ii) $l=2$ and $m=3$.
 Non-zero contributions to ${}_0\left<\beta\right|\phi_c(\mathbf{x}_c,0)\phi_B(\mathbf{x}_B,0)\left| \mathbf{\tilde{k}}^a,\mathbf{\tilde{k}}^b\right>_0$  come from
\begin{align}
{}_0\left<\beta\right| ={}_0\left<0\right| b_B(\mathbf{q}^{\bar B}) a_a(\mathbf{q}^{a})a_b(\mathbf{q}^{b})
 ~~\mathrm{or}~~
 {}_0\left<0\right|b_{\bar c}(\mathbf{q}^{\bar c})b_B(\mathbf{q}^{\bar B}) a_a(\mathbf{q}^{a})a_b(\mathbf{q}^{b})a_c(\mathbf{q}^{c}).
\end{align}
 Eq.(\ref{eq:L-S:three_body_bound_part}) thus becomes
\begin{eqnarray}
&&{}_{\rm in}\langle 0\vert \Phi_l(\mathbf{x}_2) \vert \mathbf{K}_3\rangle_0\notag \\
&=&\frac{1}{2\pi}
\frac{
T^{\dagger}_{0-ab\bar{B}}(0;{\mathbf{k}^a,\mathbf{k}^b},\mathbf{q}^{\bar B}=\mathbf{k}^c)
e^{i\mathbf{k}^c\cdot (\mathbf{x}_c-\mathbf{x}_B)}
}{(2\pi)^3
(4E_{\mathbf{k}^{c}} E^B_{\mathbf{k}^{B}})^{1/2} (-E_{\mathbf{k}^{a},\mathbf{k}^b,\mathbf{q}^{\bar B}=\mathbf{k}^c}) } \notag \\
&+&\frac{1}{2\pi}\int d^3q^{\bar c}
\frac{
T^{\dagger}_{0-abc\bar{c}\bar{B}}(0;{\mathbf{k}^a,\mathbf{k}^b},\mathbf{k}^c, \mathbf{q}^{\bar c},\mathbf{q}^{\bar B}=\mathbf{q}^{\bar c})
e^{i\mathbf{q}^{\bar c}\cdot (\mathbf{x}_B-\mathbf{x}_c)}
}{(2\pi)^3
(4E_{\mathbf{q}^{\bar c}} E^B_{\mathbf{q}^{\bar c}})^{1/2} (-E_{\mathbf{k}^{a},\mathbf{k}^b,\mathbf{k}^c,\mathbf{q}^{\bar c},\mathbf{q}^{\bar B}=\mathbf{q}^{\bar c}}) },
\end{eqnarray}
whose second term vanishes as $\vert \mathbf{x}_B-\mathbf{x}_c\vert \rightarrow \infty$. 
Thus we have
\begin{eqnarray}
{}_{\rm in}\langle 0\vert \Phi_2(\mathbf{x}_l) \vert \mathbf{K}_3\rangle_0
&\simeq& D_{23}(\mathbf{K}_3) e^{i\mathbf{k}^c\cdot (\mathbf{x}^c-\mathbf{x}_B)},\\
D_{23}(\mathbf{K}_3) &=&
\frac{1}{2\pi}
\frac{
T^{\dagger}_{0-ab\bar{B}}(0;{\mathbf{k}^a,\mathbf{k}^b},\mathbf{k}^c)
}{(2\pi)^3
(4E_{\mathbf{k}^{c}} E^B_{\mathbf{k}^{c}})^{1/2} (-E_{\mathbf{k}^{a},\mathbf{k}^b,\mathbf{q}^B=\mathbf{k}^c}) } .
\end{eqnarray}

\bibliographystyle{ptephy}
\bibliography{HALQCD}

\begin{thebibliography}{10}

\bibitem{Luscher:1990ux}
Martin Luscher, Nucl. Phys., {\bf B354}, 531--578 (1991).

\bibitem{Lin:2001ek}
C.~J.~David Lin, G.~Martinelli, Christopher~T. Sachrajda, and M.~Testa, Nucl.
  Phys., {\bf B619}, 467--498 (2001),  {{arXiv:hep-lat/0104006}}.

\bibitem{Aoki:2005uf}
S.~Aoki et~al., Phys. Rev., {\bf D71}, 094504 (2005),
  {{arXiv:hep-lat/0503025}}.

\bibitem{Ishizuka:2009bx}
N.~Ishizuka, PoS, {\bf LAT2009}, 119 (2009),  {{arXiv:0910.2772}}.

\bibitem{Ishii:2006ec}
N.~Ishii, S.~Aoki, and T.~Hatsuda, Phys. Rev. Lett., {\bf 99}, 022001 (2007),
  {{arXiv:nucl-th/0611096}}.

\bibitem{Aoki:2009ji}
Sinya Aoki, Tetsuo Hatsuda, and Noriyoshi Ishii, Prog. Theor. Phys., {\bf 123},
  89--128 (2010),  {{arXiv:0909.5585}}.

\bibitem{Aoki:2012tk}
Sinya Aoki, Takumi Doi, Tetsuo Hatsuda, Yoichi Ikeda, Takashi Inoue, Noriyoshi
  Ishii, Keiko Murano, Hidekatsu Nemura, and Kenji Sasaki, PTEP, {\bf 2012},
  01A105 (2012),  {{arXiv:1206.5088}}.

\bibitem{Carbonell:2016ekx}
J.~Carbonell and V.~A. Karmanov, Phys. Lett., {\bf B754}, 270--274 (2016),
  {{arXiv:1601.00297}}.

\bibitem{Aoki:2013cra}
Sinya Aoki, Noriyoshi Ishii, Takumi Doi, Yoichi Ikeda, and Takashi Inoue, Phys.
  Rev., {\bf D88}(1), 014036 (2013),  {{arXiv:1303.2210}}.

\bibitem{Iritani:2016jie}
Takumi Iritani et~al., JHEP, {\bf 10}, 101 (2016),  {{arXiv:1607.06371}}.

\bibitem{Iritani:2017rlk}
Takumi Iritani, Sinya Aoki, Takumi Doi, Testuo Hatsuda, Yoichi Ikeda, Takashi
  Inoue, Noriyoshi Ishii, Hidekatsu Nemura, and Kenji Sasaki, Phys. Rev., {\bf
  D96}(3), 034521 (2017),  {{arXiv:1703.07210}}.

\bibitem{Iritani:2018zbt}
Takumi Iritani, Sinya Aoki, Takumi Doi, Shinya Gongyo, Tetsuo Hatsuda, Yoichi
  Ikeda, Takashi Inoue, Noriyoshi Ishii, Hidekatsu Nemura, and Kenji Sasaki
  (2018),  {{arXiv:1805.02365}}.

\bibitem{Nemura:2008sp}
Hidekatsu Nemura, Noriyoshi Ishii, Sinya Aoki, and Tetsuo Hatsuda, Phys. Lett.,
  {\bf B673}, 136--141 (2009),  {{arXiv:0806.1094}}.

\bibitem{Inoue:2010hs}
Takashi Inoue, Noriyoshi Ishii, Sinya Aoki, Takumi Doi, Tetsuo Hatsuda, Yoichi
  Ikeda, Keiko Murano, Hidekatsu Nemura, and Kenji Sasaki, Prog. Theor. Phys.,
  {\bf 124}, 591--603 (2010),  {{arXiv:1007.3559}}.

\bibitem{Inoue:2010es}
Takashi Inoue, Noriyoshi Ishii, Sinya Aoki, Takumi Doi, Tetsuo Hatsuda, Yoichi
  Ikeda, Keiko Murano, Hidekatsu Nemura, and Kenji Sasaki, Phys. Rev. Lett.,
  {\bf 106}, 162002 (2011),  {{arXiv:1012.5928}}.

\bibitem{Murano:2011nz}
Keiko Murano, Noriyoshi Ishii, Sinya Aoki, and Tetsuo Hatsuda, Prog. Theor.
  Phys., {\bf 125}, 1225--1240 (2011),  {{arXiv:1103.0619}}.

\bibitem{Doi:2011gq}
Takumi Doi, Sinya Aoki, Tetsuo Hatsuda, Yoichi Ikeda, Takashi Inoue, Noriyoshi
  Ishii, Keiko Murano, Hidekatsu Nemura, and Kenji Sasaki, Prog. Theor. Phys.,
  {\bf 127}, 723--738 (2012),  {{arXiv:1106.2276}}.

\bibitem{Inoue:2011ai}
Takashi Inoue, Sinya Aoki, Takumi Doi, Tetsuo Hatsuda, Yoichi Ikeda, Noriyoshi
  Ishii, Keiko Murano, Hidekatsu Nemura, and Kanji Sasaki, Nucl. Phys., {\bf
  A881}, 28--43 (2012),  {{arXiv:1112.5926}}.

\bibitem{Murano:2013xxa}
K.~Murano, N.~Ishii, S.~Aoki, T.~Doi, T.~Hatsuda, Y.~Ikeda, T.~Inoue,
  H.~Nemura, and K.~Sasaki, Phys. Lett., {\bf B735}, 19--24 (2014),
  {{arXiv:1305.2293}}.

\bibitem{Kurth:2013tua}
T.~Kurth, N.~Ishii, T.~Doi, S.~Aoki, and T.~Hatsuda, JHEP, {\bf 12}, 015
  (2013),  {{arXiv:1305.4462}}.

\bibitem{Ikeda:2013vwa}
Yoichi Ikeda, Bruno Charron, Sinya Aoki, Takumi Doi, Tetsuo Hatsuda, Takashi
  Inoue, Noriyoshi Ishii, Keiko Murano, Hidekatsu Nemura, and Kenji Sasaki,
  Phys. Lett., {\bf B729}, 85--90 (2014),  {{arXiv:1311.6214}}.

\bibitem{Etminan:2014tya}
Faisal Etminan, Hidekatsu Nemura, Sinya Aoki, Takumi Doi, Tetsuo Hatsuda,
  Yoichi Ikeda, Takashi Inoue, Noriyoshi Ishii, Keiko Murano, and Kenji Sasaki,
  Nucl. Phys., {\bf A928}, 89--98 (2014),  {{arXiv:1403.7284}}.

\bibitem{Yamada:2015cra}
Masanori Yamada, Kenji Sasaki, Sinya Aoki, Takumi Doi, Tetsuo Hatsuda, Yoichi
  Ikeda, Takashi Inoue, Noriyoshi Ishii, Keiko Murano, and Hidekatsu Nemura,
  PTEP, {\bf 2015}(7), 071B01 (2015),  {{arXiv:1503.03189}}.

\bibitem{Sasaki:2015ifa}
Kenji Sasaki, Sinya Aoki, Takumi Doi, Tetsuo Hatsuda, Yoichi Ikeda, Takashi
  Inoue, Noriyoshi Ishii, and Keiko Murano, PTEP, {\bf 2015}(11), 113B01
  (2015),  {{arXiv:1504.01717}}.

\bibitem{Ikeda:2016zwx}
Yoichi Ikeda, Sinya Aoki, Takumi Doi, Shinya Gongyo, Tetsuo Hatsuda, Takashi
  Inoue, Takumi Iritani, Noriyoshi Ishii, Keiko Murano, and Kenji Sasaki, Phys.
  Rev. Lett., {\bf 117}(24), 242001 (2016),  {{arXiv:1602.03465}}.

\bibitem{Miyamoto:2017tjs}
Takaya Miyamoto et~al., Nucl. Phys., {\bf A971}, 113--129 (2018),
  {{arXiv:1710.05545}}.

\bibitem{Kawai:2017goq}
Daisuke Kawai, Sinya Aoki, Takumi Doi, Yoichi Ikeda, Takashi Inoue, Takumi
  Iritani, Noriyoshi Ishii, Takaya Miyamoto, Hidekatsu Nemura, and Kenji
  Sasaki, PTEP, {\bf 2018}(4), 043B04 (2018),  {{arXiv:1711.01883}}.

\bibitem{Ikeda:2017mee}
Yoichi Ikeda, J. Phys., {\bf G45}(2), 024002 (2018),  {{arXiv:1706.07300}}.

\bibitem{Gongyo:2017fjb}
Shinya Gongyo et~al., Phys. Rev. Lett., {\bf 120}, 212001 (2018),
  {{arXiv:1709.00654}}.

\bibitem{Sasaki:2017ysy}
Kenji Sasaki et~al., PoS, {\bf LATTICE2016}, 116 (2017),  {{arXiv:1702.06241}}.

\bibitem{Ishii:2017xud}
Noriyoshi Ishii et~al., PoS, {\bf LATTICE2016}, 127 (2017),
  {{arXiv:1702.03495}}.

\bibitem{Doi:2017cfx}
Takumi Doi et~al., PoS, {\bf LATTICE2016}, 110 (2017),  {{arXiv:1702.01600}}.

\bibitem{Nemura:2017bbw}
Hidekatsu Nemura et~al., PoS, {\bf LATTICE2016}, 101 (2017),
  {{arXiv:1702.00734}}.

\bibitem{Doi:2017zov}
Takumi Doi et~al.,
\newblock {Baryon interactions from lattice QCD with physical quark masses --
  Nuclear forces and $\Xi\Xi$ forces --},
\newblock In {\em {35th International Symposium on Lattice Field Theory
  (Lattice 2017) Granada, Spain, June 18-24, 2017}} (2017),
  {{arXiv:1711.01952}}.

\bibitem{Nemura:2017vjc}
Hidekatsu Nemura et~al.,
\newblock {Baryon interactions from lattice QCD with physical masses ---
  strangeness $S=-1$ sector ---},
\newblock In {\em {35th International Symposium on Lattice Field Theory
  (Lattice 2017) Granada, Spain, June 18-24, 2017}} (2017),
  {{arXiv:1711.07003}}.

\bibitem{Aoki:2010ry}
Sinya Aoki,
\newblock {Lattice QCD and Nuclear Physics},
\newblock In {\em {Modern perspectives in lattice QCD: Quantum field theory and
  high performance computing. Proceedings, International School, 93rd Session,
  Les Houches, France, August 3-28, 2009}}, pages 591--628 (2010),
  {{arXiv:1008.4427}}.

\bibitem{weinberg2005quantum}
Steven Weinberg,
\newblock {\em The Quantum Theory of Fields: Volume 1, Foundations},
\newblock  (Cambridge University Press, 2005).

\bibitem{HALQCD:2012aa}
Noriyoshi Ishii, Sinya Aoki, Takumi Doi, Tetsuo Hatsuda, Yoichi Ikeda, Takashi
  Inoue, Keiko Murano, Hidekatsu Nemura, and Kenji Sasaki, Phys. Lett., {\bf
  B712}, 437--441 (2012),  {{arXiv:1203.3642}}.

\end{thebibliography}
\end{document}